\documentclass[11pt]{article}
\usepackage{graphicx}

\newsavebox{\PSLASH}
\sbox{\PSLASH}{$p$\hspace{-1.8mm}/}

\begin{document}

\title{Conformal symmetry in  non-local field theories}

\author{ M. A. Rajabpour$^{a}$\footnote{e-mail: rajabpour@sissa.it} \\  \\
  $^{a}$SISSA and INFN, 
\textit{Sezione di Trieste},  via Bonomea 265, 34136 Trieste, Italy}

\maketitle
\begin{abstract}
 We have shown that  a particular class of 
non-local free  field theory  has conformal symmetry in arbitrary dimensions. Using the local field theory counterpart of  this  class,  we have found the Noether currents and  Ward identities of the translation, rotation and scale symmetries. The operator product expansion of the  energy-momentum tensor with quasi-primary fields is also investigated. 
\end{abstract}

\section{Introduction}\
Non-local field theories are well-known as a method to describe the scaling limit of the long-range interacting systems  and they are much studied in statistical physics and string theory. Long-range spin systems \cite{FNM}, rough surfaces \cite{rough} and diffusion processes \cite{Gornflo} are just few examples among many  in statistical physics.

Although many studies have dealt with  non-local field theories, mostly calculating the renormalization group equations and scaling properties,  we still know very little about the role of the symmetries in these systems. 
One of the very powerful symmetries is  conformal symmetry, which is the Poincare symmetry plus scale and special conformal symmetry. It is powerful because it puts several restrictions on the form of the correlation functions. For an extensive study of conformal field theory (CFT), see \cite{DMS,Cardy1}. For a recent  tutorial on  conformal symmetry in diverse dimensions, see \cite{JP}. Since special conformal symmetry is a symmetry made of translation and inversion,  it could be quite surprising if we  find conformal symmetry in non-local field theories. The rule for conformal symmetry in non-local field theories is already discussed in the string theory community by usually defining the theory by its correlation functions\footnote{There are also some comments about the conformal symmetry at the level of the non-local Lagrangian \cite{RV}.} \cite{HPPS, EP}. Some deformed versions of conformal symmetry in non-local field theories are also discussed in ageing systems, see \cite{Malte1,Malte2} and reference therein. In this paper we will study systematically the fractional Laplacian field theory (at the level of the lagrangian) which is a generalization of the  field theory with  different powers of the Laplacian. We first show that this field theory has conformal symmetry in any dimension. In two dimensions  it has just global conformal symmetry and not full conformal symmetry. Then we will study different consequences of conformal symmetry in these field theories. 

 The paper is organized as follows. In the  section ~2 we define the fractional Laplacian and we show that the fractional Laplacian field theory has conformal symmetry in arbitrary dimensions. In the section ~3  we will introduce a method to find the Noether currents of the symmetries in non-local field theories by mapping the non-local field theory to a local field theory. We also study the Ward identities of the translational, rotational and scale invariance. In the section ~4 we investigate the operator product expansion structure in the non-local field theories.  Finally in the last section we  will summarize our findings and  give some comments about possible future directions.  

\section{Conformal symmetry of the fractional Laplacian field theory}\
\setcounter{equation}{0}
In this section we will discuss the conformal symmetry of the fractional Laplacian field theory in general dimensions. The action of the fractional Laplacian field theory is as follows:
 \begin{eqnarray}\label{fractional laplacian field theory}
S=\frac{1}{2}\int \Phi(x) (-\bigtriangleup)^{\frac{\alpha}{2}}\Phi(x)d^dx,
\end{eqnarray}
where $(-\bigtriangleup)^{\frac{\alpha}{2}}$ is the fractional laplacian defined by its Fourier transform
\begin{eqnarray}\label{fractional laplacian definition1}
\widehat{(-\bigtriangleup)^{\frac{\alpha}{2}}}\Phi(\mathbf{p})=|\mathbf{p}|^\alpha \widehat{\Phi(\mathbf{p})}.
\end{eqnarray}
Here we define the Fourier transform as 
\begin{eqnarray}\label{Fourier transform}
F[f(x)]=\widehat{f(\mathbf{p})}=\int f(\mathbf{x})e^{i\mathbf{p}.\mathbf{x}}d^d\mathbf{x}.
\end{eqnarray}
The non-local nature of the fractional Laplacian can be seen in its representation in the real space, which is
\begin{eqnarray}\label{fractional laplacian definition2}
(-\bigtriangleup)^{\frac{\alpha}{2}}\Phi(x)=C(d,\alpha)\int \frac{\Phi(x)-\Phi(y)}{|x-y|^{d+\alpha}}d^dy,\hspace{1cm}0 < \alpha < 2,
\end{eqnarray}
where $C(d,\alpha)$ is a constant \cite{Samko}. The inverse of the  fractional Laplacian is the Riesz potential and has the following form
\begin{eqnarray}\label{Riesz potential}
(-\bigtriangleup)^{-\frac{\alpha}{2}}\Psi(x)=D_1(d,\alpha)\int \frac{\Psi(y)}{|x-y|^{d-\alpha}}d^dy,\hspace{1cm}\hspace{1cm}0 < \alpha < d,
\end{eqnarray}
where $\Psi(x)=(-\bigtriangleup)^{\frac{\alpha}{2}}\Phi(x)$ and $D_1(d,\alpha)=\frac{\Gamma(\frac{d-\alpha}{2})}{2^\alpha \pi^{\frac{d}{2}}\Gamma(\frac{\alpha}{2})}$.

The conformal algebra in $d$ dimensions is made of $d$ generators of  translation, $\frac{d(d-1)}{2}$ generators of  rotation, $1$ generator of the scale transformation and $d$ generators of the special conformal transformation. It is quite simple to see that the action (\ref{fractional laplacian field theory}) has  translational, rotational and scale symmetry for $0 < \alpha < 2$. The invariance of the action with respect to the special conformal transformation, which is composed of translation plus  inversion, is not trivial. The transformation corresponding to  inversion is $\mathbf{x}\rightarrow{\frac{\mathbf{x}}{|\mathbf{x}|^2}}$. To see that the  action (\ref{fractional laplacian field theory}) is conformally invariant,  we need to show that under the  transformation 
\begin{eqnarray}\label{conformal transform}
\mathbf{x}\rightarrow{\mathbf{x}'=g(\mathbf{x})},\hspace{1cm}\Phi'(\mathbf{x}')=|J_g|^{\frac{-d+\alpha}{2}}\Phi(\mathbf{x}),
\end{eqnarray}
where $g(\mathbf{x})$ is the conformal transformation and $|J_g|$ is the $d$th root of the Jacobian of the transformation,  the action does not change. To have conformal symmetry, we need to show that
\begin{equation}\label{conformal transform1}
\frac{1}{2}\int \Phi(x) (-\bigtriangleup)^{\frac{\alpha}{2}}\Phi(x)d^dx=\frac{1}{2}\int \Phi'(x') (-\bigtriangleup')^{\frac{\alpha}{2}}\Phi'(x')d^dx',
\end{equation}
or
\begin{equation}\label{conformal transform2}
|J_g|^{\frac{\alpha+d}{2}}(-\bigtriangleup)^{\frac{\alpha}{2}} \Phi'(x)|_{x=x'} =(-\bigtriangleup)^{\frac{\alpha}{2}}(|J_g|^{\frac{-\alpha+d}{2}}\Phi'(x')).
\end{equation}

Using the Jacobian of the the inversion transformation, i.e. $|J|=\frac{1}{|\mathbf{x}|^2}$, and using the  equality
\begin{eqnarray}\label{equality}
|\mathbf{x}|^{-\alpha-d}\int |\frac{\mathbf{x}}{|\mathbf{x}|^2}-\mathbf{y}|^{-\alpha-d}f(\mathbf{y})d^dy=\int |\mathbf{x}-\mathbf{y}|^{-\alpha-d}|\mathbf{y}|^{-d+\alpha}f(\frac{\mathbf{y}}{|\mathbf{y}|^2})d^dy,
\end{eqnarray}
one can easily show that  equation (\ref{conformal transform2}) is valid for the inversion transformation. Checking the validity of  equality (\ref{conformal transform1}) for  translation, rotation and scale transformations is almost a trivial task. 
The above calculation shows that the action (\ref{fractional laplacian field theory}) is conformally invariant for $0 < \alpha < d$. The definition of the case $d=\alpha$ is very similar to the previous one, we just need to consider logarithm in the definition of the Riesz potential 
\begin{eqnarray}\label{Riesz potential2}
(-\bigtriangleup)^{-\frac{d}{2}}\Psi(x)=D_2(d,\alpha)\int \log|\mathbf{x}-\mathbf{y}|^{-1}\Psi(y)d^dy,
\end{eqnarray}
where $D_2(d,\alpha)=\frac{1}{(4\pi)^{\frac{d}{2}}\Gamma(\frac{d}{2})}$. The ordinary Laplacian is just $\alpha=d=2$, and it is well-known that it is fully conformally invariant. In the appendix we show that $\alpha=d=4$ is also conformally invariant. We conjecture that the action (\ref{fractional laplacian field theory}) is conformally invariant for all positive real $\alpha$'s. The field $\Phi(\mathbf{x})$ is a quasi-primary field with conformal weight $x_{\phi}=\frac{d-\alpha}{2}$; in other words, the two-point correlation function  of the $\Phi(\mathbf{x})$ operator is
\begin{eqnarray}\label{correlation function}
<\Phi(x)\Phi(0)> =
\left\{
\begin{array}{lr}
D_1(d,\alpha)\frac{1}{|x|^{d-\alpha}},&0 < \alpha < d,\\
D_2(d,\alpha) \log|x|^{-1},&\alpha = d.
\end{array}
\right.
\end{eqnarray}
The operator $\Phi(x)$ is not the only operator with power law correlations, for  $\alpha = d$ it is easy to see that the operator $e^{i\beta\Phi(x)}$ is also quasi-primary and 
\begin{eqnarray}\label{correlation function vertex oparator}
<e^{i\beta\Phi(x)}e^{-i\beta\Phi(0)}>=\frac{1}{|x|^{\beta^2 D_2(d,\alpha)}}.
\end{eqnarray}
It is not difficult to see that to get a non-trivial correlation function one needs to fix the charges to zero, in other words
\begin{eqnarray}\label{correlation function vertex oparators}
<e^{i\alpha_1\Phi(x_1)}e^{i\alpha_2\Phi(x_2)}...e^{i\alpha_n\Phi(x_n)}>=0,
\end{eqnarray}
if $\alpha_1+\alpha_2+...+\alpha_n \neq 0$. This is a consequence of the symmetry of the action with respect to $\Phi(x)\rightarrow{\Phi(x)+a}$  ( this can be seen after partial integration). It is worth mentioning that the following action is also conformally invariant
\begin{eqnarray}\label{conformally invariant action}
S=\frac{1}{2}\int \Big{(}\Phi(x) (-\bigtriangleup)^{\frac{\alpha}{2}}\Phi(x)+(\Phi^2(x))^{\frac{d}{d-\alpha}}\Big{)}d^dx.
\end{eqnarray}
The scale invariance of the above action can be checked by simple dimensional argument.

\section{Ward identities}\
\setcounter{equation}{0}

The   field theory defined by (\ref{fractional laplacian field theory}) is a non-local field theory. The non-locality makes the study of the variations of the field theory difficult. One way to overcome this difficulty is to use the local counterpart of the fractional Laplacian. This can be done by following the Caffarelli and Silvestre trick \cite{CS}, see also \cite{Chang}.  The correspondence is based on the equivalence of the fractional field theory in $d$ dimensions with an ordinary Laplacian in $n=d+1$ dimensions, in other words 
\begin{eqnarray}\label{local-field-theory}
S=\frac{1}{2}\int_{y>0}  (\partial_\mu \tilde{\Phi}(x,y))^2y^{1-\alpha}d^dxdy=C\frac{1}{2}\int \Big{(}\Phi(x) (-\bigtriangleup)^{\frac{\alpha}{2}}\Phi(x)\Big{)}d^dx,
\end{eqnarray}
where $C$ is a constant and $0<\alpha<2$ (for an extension to the range $0<\alpha<d$ as long as $\alpha$ is not an integer see \cite{Chang}) and $\tilde{\Phi}(x,0)=\Phi(x)$. The above equality can be easily shown  in momentum space \cite{CS} . It is also easy to show that $\tilde{\Phi}(x,y)$ and $\Phi(x)$ have the same two point functions in  $\mathbf{x}$ space. Using the Euler-Lagrange equation of the local field theory, i.e. $\partial^\mu(y^{1-\alpha}\partial_\mu\tilde{\Phi}(x,y))=0$, one can show that 
\begin{eqnarray}\label{local-nonlocal equality}
C(-\bigtriangleup)^{\frac{\alpha}{2}}\Phi(x)=\lim_{y\rightarrow{0}}y^{1-\alpha}\tilde{\Phi}_{y}(x,y)=\delta^{d}(x).
\end{eqnarray}
Motivated by the above correspondence, we study the symmetries of the local field theory instead of the non-local fractional Laplacian. First of all we notice that our field theory is translational invariant in  $\mathbf{x}$ space so we will have $d$ generators of translation. In addition we have $\frac{d(d-1)}{2}$ generators of rotation in  $\mathbf{x}$ space and also $d$ generators of  special conformal transformations and a generator of the scale transformation. The space of the symmetries of the two field theories in the equation (\ref{local-field-theory}) is the same. The very tricky symmetry is the scale invariance which is highly dependent on the $y$ coordinate. 

The conserved currents $J^{\mu}_{a}$ of the local field theory are defined by the response of the action to the infinitesimal transformation of the coordinates $x'^{\mu}=x^{\mu}+\omega_{a}\frac{\delta x'^{\mu}}{\delta\omega_{a}}(x)$ and fields $\tilde{\Phi'}(x',y')=\tilde{\Phi}(x,y)+\omega^{a}\frac{\delta \mathcal{F}}{\delta\omega_{a}}(x,y)$. For the generic field theory $S=\int \mathcal{L}[\tilde{\Phi}(x,y),\partial_{\mu}\tilde{\Phi}(x,y)]y^{1-\alpha}d^dxdy$ the transformation of the action is 
\begin{eqnarray}\label{action-change}
S'-S=\int_{y>0} \partial_{\mu}J^{\mu}_{a}\omega^{a}d^dxdy,
\end{eqnarray}
where
\begin{eqnarray}\label{current}
J^{\mu}_{a}=y^{1-\alpha}\Big{(}(\delta_{\nu}^{\mu}\mathcal{L}-\frac{\partial\mathcal{L}}{\partial(\partial_{\mu}\tilde{\Phi}(x,y))}
\partial_{\nu}\tilde{\Phi}(x,y))\frac{\delta x^{\nu}}{\delta\omega_{a}} + \frac{\partial\mathcal{L}}{\partial(\partial_{\mu}\tilde{\Phi}(x,y))}\frac{\delta \mathcal{F}}{\delta\omega_{a}}\Big{)},
\end{eqnarray}
and $\mu=1,2,...,n$. From now on we will use italic letters for the quantities in the $\mathbf{x}$ space and the Greek letters for the quantities in the $(\mathbf{x},y)$ space. It is worth mentioning that to get the equation (\ref{action-change}) we do not need to do any partial integration. The conserved current associated with the translational invariance is the canonical energy-momentum tensor and has the following form:
\begin{eqnarray}\label{energy-momentum}
T^{c}_{\mu i}=y^{1-\alpha}\Big{(}\partial_{\mu}\tilde{\Phi}(x,y)\partial_{i}\tilde{\Phi}(x,y)-\frac{1}{2}\delta_{\mu i}(\partial\tilde{\Phi}(x,y))^2\Big{)},
\end{eqnarray}
where $i=1,2,...,d$. Using the equations of motion it is easy to show that $\partial^{\mu}T^{c}_{\mu i}=0$. Formally one can also define  $T^{c}_{\mu y}$ by the same formula, but one should keep in mind that it is not a conserved current. Since the above energy-momentum tensor is already symmetric, one can write the associated conserved current of the rotational invariance as
\begin{eqnarray}\label{current-rotation}
j_{\mu ij}=T^{c}_{\mu i}x_{j}-T^{c}_{\mu j}x_{i}.
\end{eqnarray}
 Using the equations of motion, one can again easily show that $\partial^{\mu}j_{\mu ij}=0$. The last and the most tricky conserved current is the dilation current, it can be written as 
\begin{eqnarray}\label{current-dilation}
D_{\mu}=-x_{\nu}T_{\mu}^{c\nu}+y^{1-\alpha}x_{\phi} \tilde{\Phi}(x,y)\partial_{\mu}\tilde{\Phi}(x,y),
\end{eqnarray}
where $x_{\phi}=\frac{n-1-\alpha}{2}=\frac{d-\alpha}{2}$ and $\partial^{\mu}D_{\mu}=0$. Notice that the  energy-momentum tensor is living in $d+1$ dimensions, so we expect that the weight of this energy-momentum tensor will be $d+1$. On the other hand, the non-local field theory is living in $d$ dimensions so to get a sensible energy-momentum tensor for our system it is reasonable to integrate $T(x,y)$ over $y$ from zero to infinity. We define 
\begin{eqnarray}\label{energy-momentum tensor non-local}
t_{\mu\nu}=\int_{0}^{\infty}T^{c}_{\mu\nu}dy.
\end{eqnarray}
Using the above definition, one can write the Ward identity of the translational invariance in  $\mathbf{x}$ space as
\begin{eqnarray}\label{ward identity-translation}
\partial^{j}<t_{ij}\textbf{X}>=-\sum_k \delta(x-x_k)\frac{\partial}{\partial x_{k}^{i}}<\textbf{X}>,
\end{eqnarray}
where $\textbf{X}$ is any operator living in the $\textbf{x}$ space and without any derivative over $y$. The Ward identity of the rotational symmetry is 
\begin{eqnarray}\label{ward identity-rotation}
<(t_{ij}-t_{ji})\textbf{X}>=-i\sum_k \delta(x-x_k)S_{ij}^{k}<\textbf{X}>,
\end{eqnarray}
where $S_{ij}^{k}$ is the spin generator appropriate for the $k$-th field of the set $X$. The Ward identity of the scale symmetry has a form that is a bit disappointing; 
\begin{eqnarray}\label{ward identity-scale}
<t_{i}^iX>&+&\int dy y<\partial^iT_{iy}\textbf{X}>+x_{\phi_k}\int dy y^{1-\alpha}<\partial^i\Big{(}\tilde{\Phi}(x,y)\partial_i\tilde{\Phi}(x,y)\Big{)}\textbf{X} >\nonumber\\&=&-\sum_k \delta(x-x_k)x_{\phi_k}<\textbf{X}>,
\end{eqnarray}
where $x_{\phi_k}$ is the weight appropriate for the $k$-th field of the set $\textbf{X}$. In usual CFT studies the second and the third terms in the left hand side disappear because there one starts with a traceless energy-momentum tensor which helps to write the conserved dilation current as $D_{\mu}=x^\nu T_{\nu\mu}$. It seems that this is not a trivial task to do in our model because making $T_{\mu\nu}$ traceless in $n$ dimensions does not mean that we can write $D_{\mu}$ fully with respect to the energy-momentum tensor. The trouble comes from the fact that the energy-momentum tensor is not conserved in the $y$ direction while $D_{\mu}$ is fully affected by also the $y$ space. In other words $\partial^\mu D_\mu=T^{\mu}_{\mu}+x^{\nu}\partial^\mu T_{\nu\mu}=T_\mu^\mu+y\partial^\mu T_{\mu y}$, where the second term is zero in the usual CFT due to the conservation of the energy-momentum tensor in the whole space. Another way to view this problem is to look at the form of the virial in the equation (\ref{ward identity-scale}), $V_i=\int dy yT_{iy}+x_{\phi}\int dy y^{1-\alpha}\Big{(}\tilde{\Phi}(x,y)\partial_i\tilde{\Phi}(x,y)\Big{)}$. Although it is possible to write the second term as the gradient of a tensor,  it is impossible to do the same with the first term. This is in contrast with what one expects from the previous arguments \cite{JP,Polchinski} which state that  the virial should be a gradiant of a tensor for  conformally invariant models (for an example see the appendix). This means that although $t_{ij}$ carries some properties of the energy-momentum tensor in the $\textbf{x}$ space it is not exactly the same as the conventional energy-momentum tensor.  
In two dimensions, it is probably impossible to find a traceless energy-momentum tensor because tracelessness means that the action is invariant under full conformal symmetry \cite{Polchinski} which we already know  is not true in our case.

\section{Operator product expansion}\
\setcounter{equation}{0}

In this section, we study the leading terms of the operator product expansion of the  energy-momentum tensor with the primary fields.  For simplicity we first focus on the  case $\alpha=1$. Although the explicit dependence on $y$ in the action and the energy-momentum tensor disappears, the original field theory is still a non-local field theory.  For this case one can simply use the traceless energy-momentum tensor of the bosonic field theory in $n$ dimensions. To bring out the connection to  well-known facts in bosonic field theory, we write the action as
\begin{eqnarray}\label{bosonic action}
S=\frac{1}{2}\int (\partial_\mu \tilde{\Phi}(x,y))^2d^dxdy,
\end{eqnarray}
with the following two point function
\begin{eqnarray}\label{two point function}
<\tilde{\Phi}(x,y)\tilde{\Phi}(0,0)>=\frac{1}{S_n(n-2)r^{n-2}},
\end{eqnarray}
 where $r=\sqrt{\mathbf{x}^2+y^2}$ and  $S_n=\frac{2\pi^{n/2}}{\Gamma[\frac{n}{2}]}$. We define the energy-momentum tensor with the following equation\footnote{The normalization is chosen to be coherent with the well-known results for $\alpha=1$, see for example \cite{Capelli}. As we will see for $\alpha\neq 1$ different normalization leads to the known OPEs.}:
\begin{eqnarray}\label{renormalized energy-momentum tensor}
\delta S=-\frac{1}{S_n}\int \partial^\mu T_{\mu\nu}\omega^\nu d^dxdy.
\end{eqnarray}
 Then the traceless energy-momentum tensor is  
\begin{eqnarray}\label{energy-momentum2}
\frac{T_{\mu \nu}}{S_n}=-\frac{1}{2(n-1)}\Big{(}n\partial_{\mu}\tilde{\Phi}(x,y)\partial_{\nu}\tilde{\Phi}(x,y)-(n-2)\tilde{\Phi}(x,y)\partial_{\mu}\partial_{\nu}\tilde{\Phi}(x,y)-\nonumber\\\delta_{\mu \nu}\Big{(}(\partial\tilde{\Phi}(x,y))^2-\frac{n-2}{n}\tilde{\Phi}(x,y)\partial^2\tilde{\Phi}(x,y)\Big{)}\Big{)}.
\end{eqnarray}
 The operator product expansion of the above energy-momentum tensor with a primary operator is\footnote{This form is valid for generic CFT with generic primary operators \cite{Cardy2}.} 
\begin{eqnarray}\label{OPEbosonic-alpha=1}
T_{\mu\nu}\tilde{\Phi}(0,0)=A_{\mu\nu}^n\tilde{\Phi}(0,0)+B_{\mu\nu\lambda}^n\partial^{\lambda}\tilde{\Phi}(0,0),
\end{eqnarray}
where
\begin{eqnarray}\label{A and B}
A_{\mu\nu}^n&=&\frac{\frac{nx_{\tilde{\phi}}}{n-1} (r_\mu r_\nu-\frac{1}{n}r^2\delta_{\mu\nu})}{r^{n+2}},\\
B_{\mu\nu\lambda}^n&=&B_{\mu\nu\lambda}^{1n}+B_{\mu\nu\lambda}^{2n},\\
B_{\mu\nu\lambda}^{1n}&=&\frac{n}{2(n-1)}\Big{(}\frac{r_\mu\delta_{\nu\lambda}+r_\nu\delta_{\mu\lambda}-(2/n)r_\lambda \delta_{\mu\nu}}{r^n}\Big{)},\\
B_{\mu\nu\lambda}^{2n}&=&
\frac{n(n-2)}{2(n-1)}\Big{(}\frac{(r_\mu r_\nu-\frac{1}{n}r^2\delta_{\mu\nu})r_\lambda}{r^{n+2}}\Big{)},
\end{eqnarray}
with $x_{\tilde{\phi}}=\frac{n-2}{2}$. The integration over $y$ gives
\begin{eqnarray}\label{OPE-alpha=1 }
t_{ij}\Phi(x)=\frac{\sqrt{\pi}d\Gamma(\frac{d}{2})}{4\Gamma(\frac{d+3}{2})}A_{ij}^d\Phi(0)+\Big{(}\frac{\sqrt{\pi}d\Gamma(\frac{d}{2})}{2\Gamma(\frac{d+1}{2})}B_{ijk}^{1d}+d(d-1)\frac{\sqrt{\pi}\Gamma(\frac{d}{2})}{4\Gamma(\frac{d+3}{2})}B_{ijk}^{2d}\Big{)}\partial^{k}\Phi(0)+...,
\end{eqnarray}
where 

\begin{eqnarray}\label{A and B after integration}
A_{ij}^d&=&\frac{\frac{(d+1)x_{\phi}}{d} (r_i r_j-\frac{1}{d}r^2\delta_{ij})}{r^{d+2}},\\
B_{ijk}^{1d}&=&\frac{d+1}{2d}\Big{(}\frac{r_i\delta_{jk}+r_j\delta_{ik}-(2/d)r_k \delta_{ij}}{r^d}\Big{)}\\
B_{ijk}^{2d}&=&
\frac{(d-1)(d+1)}{2d}\Big{(}\frac{(r_i r_j-\frac{1}{d}r^2\delta_{ij})r_k}{r^{d+2}}\Big{)}.
\end{eqnarray}
To get an idea about the above results we now investigate the most interesting dimension $d=2$. Define the following energy-momentum tensors in the complex plane  
\begin{eqnarray}\label{energy-momentum complex plane}
t_{zz}&=&t_{00}-2it_{10}-t_{11},\\
t_{\bar{z}\bar{z}}&=&t_{00}+2it_{10}-t_{11},\\
t_{\bar{z}z}&=&\frac{1}{4}(t_{00}+t_{11}).
\end{eqnarray}
It is easy to show that
\begin{eqnarray}\label{OPE energy-momentum complex plane}
t_{zz}\Phi(0)&=&\frac{x_\phi}{z^2}\Phi(0)+\frac{5\partial_z\Phi(0,0)}{2z}+\frac{\bar{z}\partial_{\bar{z}}\Phi(0,0)}{2z^2}+...,\\
t_{\bar{z}\bar{z}}\Phi(0)&=&\frac{x_{\phi}}{\bar{z}^2}\Phi(0)+\frac{5\partial_{\bar{z}}\Phi(0,0)}{2\bar{z}}+\frac{z\partial_z\Phi(0,0)}{2\bar{z}^2}+...,
\end{eqnarray}
where $x_\phi=\frac{d-1}{2}$. All of the terms, except the most singular one, do not have the usual CFT form (in usual CFT we have $t_{zz}\Phi(0)=\frac{x_\phi}{z^2}\Phi(0)+\frac{\partial_z\Phi(0)}{z}$). This is not surprising because as we already noticed the Ward identity of the dilation symmetry does not have the usual form. The first term is very similar to the usual CFT term, however, one should be careful that the above equation still needs to be renormalized by $C$.  
The equation (\ref{OPE energy-momentum complex plane}) shows that $t_{ij}$ has some properties that we expect for the conformal field theories.  $t_{ij}$  carries some important features of the energy-momentum tensor; for example, (\ref{OPE energy-momentum complex plane}) suggests that one can consider $t_{zz}(t_{\bar{z}\bar{z}})$ as a holomorphic (anti-holomorphic) operator with spin $2(-2)$. Getting the OPE of the energy-momentum tensor with itself is a non-trivial task. Doing blindly the two integrations over the $y$ space of the OPE of the two energy-momentum tensors in the (\textbf{x},y) space will give the infinity. This seems consistent with the observation in \cite{EP}, however, we think that more study in this direction is necessary.  

For generic $\alpha$ one can consider the following conserved energy-momentum tensor
\begin{eqnarray}\label{energy-momentum3}
\frac{T_{\mu i}}{S_n}=-\frac{y^{1-\alpha}}{2(n-1)}\Big{(}n\partial_{\mu}\tilde{\Phi}(x,y)\partial_{i}\tilde{\Phi}(x,y)-(n-2)\tilde{\Phi}(x,y)\partial_{\mu}\partial_{i}\tilde{\Phi}(x,y)-\delta_{\mu i}(\partial\tilde{\Phi}(x,y))^2\Big{)},
\end{eqnarray}
where $\partial^\mu T_{\mu i}=0$. Note that the above energy-momentum tensor is symmetric but not traceless, i.e. $T^i_i\neq 0$.  Then the OPE of $T$ with $\tilde{\Phi}$ after using (\ref{two point function}) and Wick's theorem is
\begin{eqnarray}\label{OPE energy-momentum 2}
T_{\mu \nu}\tilde{\Phi}(0,0)=\frac{(n+1-\alpha)y^{1-\alpha}}{2(n-1)}\frac{ (r_\mu r_\nu-\frac{1}{n+1-\alpha}r^2\delta_{\mu\nu})}{r^{n+3-\alpha}}\tilde{\Phi}(0,0)+....
\end{eqnarray}
After integration over $y$ one can simply write
\begin{eqnarray}\label{OPE energy-momentum 2 after integration}
t_{ij}\Phi(0)=\frac{(d-1)\Gamma(\frac{d}{2})\Gamma(1-\frac{\alpha}{2})}{4\Gamma(\frac{2+d-\alpha}{2})}\frac{ (r_i r_j-\frac{1}{d}r^2\delta_{ij})}{r^{d+2}}\Phi(0)+....
\end{eqnarray}
Renormalization of $t_{ij}$ will lead us to the well-known form of the first term of the  OPE of the energy-momentum tensor and a primary field.

\section{Conclusion}\
In this paper we showed that non-local field theories can have conformal symmetry in arbitrary dimensions. This gives an example of a field theory in two dimensions which is globally conformally invariant but not fully conformally
symmetric \cite{Polchinski,Riva,slava2}. Using the local counterpart of the non-local field theory, we proposed a way to extract the Noether currents of the non-local field theories. The Ward identities of the translation, rotation and scale transformations were derived. The Ward identity of the  scale symmetry does not have the  usual form that we expect in CFT. In other words we have not been able to write all the conserved currents with respect to the energy-momentum tensor. We do not have a particular argument to rule out the possibility of writing all the conserved currents with respect to an improved energy-momentum tensor except in  two dimensions. Finding such an improved energy-momentum-tensor is a subject for further studies in this direction. We should point out here that since our starting point was the fractional Laplacian field theory, we encountered the possibility of using the fractional derivative in the construction of the operator structure of our model. This is the reason that we have energy-momentum tensor in our construction which is absent in the earlier discussions \cite{HPPS}.

 Using the energy-momentum tensor defined in the third section, we also calculated  the different operator product expansions. Some terms in the operator product expansions (the most singular terms) have forms very similar to the cases in the usual CFTs, the less singular terms are quite different from the usual CFT. We have not been able to find a consistent OPE for the two energy-momentum tensors. 

\textbf{Acknowledgments:}
The author is indebted to Sebastian Guttenberg for many useful comments and discussions. I also thank Malte Henkel for discussions and introducing me to the book \cite{Malte2} and Slava Rychkov for drawing my attention to the references \cite{RV,HPPS,EP,slava2}. I also thank R. Jackiw, S. Rouhani and R. T. Requist for reading the manuscript.

\section*{Appendix :Symmetric traceless energy-momentum tensor of the fourth derivative
free theory
}
In this appendix we study the symmetries of the fourth derivative free theory. The purpose of this appendix is to give an idea about the difference between the  conformal symmetry in the two dimensions and  higher dimensions. We show that it is possible to find a symmetric traceless energy-momentum tensor in $d>2$ but it is impossible to find a traceless-energy-momentum tensor in $d=2$ except for $\alpha=2$. The action of the fourth derivative free theory is
\begin{eqnarray}\label{fourth derivative scalar theory}
S=\int d^dx (\bigtriangleup\psi(x))^2,
\end{eqnarray}
where $\bigtriangleup\psi(x)=\partial_\mu\partial^\mu\psi(x)$. Consider the transformation (\ref{conformal transform}) with $\alpha=4$. The canonical energy-momentum tensor which is the generator of translation  has the following form
\begin{eqnarray}\label{canonical energy-momentum fourth derivative scalar theory}
T^c_{\mu\nu}=-2\partial_\mu\partial_\nu\psi(x)\bigtriangleup\psi(x)+2\partial_\nu\psi\partial_\mu(\bigtriangleup\psi(x))+\delta_{\mu\nu}(\bigtriangleup\psi(x))^2.
\end{eqnarray}
 The above energy-momentum tensor is not symmetric. To get a symmetric energy-momentum tensor one can use the following Belinfante tensor defined for arbitrary dimension:

\begin{eqnarray}\label{Beleinfante}
B_{\rho\mu\nu}=-2\delta_{\mu\nu}\bigtriangleup\psi(x)\partial_{\rho}\psi+2\delta_{\rho\nu}\bigtriangleup\psi(x)\partial_{\mu}\psi,
\end{eqnarray}
which is antisymmetric with respect to the first two indices. Using the above tensor one can write the Belinfante energy-momentum tensor as 
\begin{eqnarray}\label{Beleinfante energy-momentum}
T^B_{\mu\nu}=T^c_{\mu\nu}+\partial^\rho B_{\mu\rho\nu}=2\partial_\nu\psi\partial_\mu(\bigtriangleup\psi(x))+\hspace{3cm}\nonumber\\\hspace{2cm}2\partial_\mu\psi\partial_\nu(\bigtriangleup\psi(x))-\delta_{\mu\nu}\Big{(}(\bigtriangleup\psi(x))^2+2\partial_\rho\psi\partial_{\rho}(\bigtriangleup\psi(x))\Big{)}.
\end{eqnarray}
Using the above energy-momentum tensor, one can simply write the generator of the rotation as 
\begin{eqnarray}\label{rotation}
j_{\mu\nu\rho}= T^B_{\mu\nu}x_\rho-T^B_{\mu\rho}x_\nu.
\end{eqnarray}
The generator of the scale invariance is 
\begin{eqnarray}\label{scale}
D^\mu=x^\nu T^{c\mu}_\nu-\frac{d-2}{2}\bigtriangleup\psi(x)\partial^\mu\psi+(\frac{d}{2}-3)\psi\partial^\mu \bigtriangleup\psi(x).
\end{eqnarray}
We define now the following tensor 
\begin{eqnarray}\label{sigma}
\sigma^{\alpha\mu}=2\partial^\alpha\psi\partial^\mu\psi-2\psi\partial^\alpha\partial^\mu\psi+(d-2)\delta^{\alpha\mu}\psi \bigtriangleup\psi(x),
\end{eqnarray}
then the virial of the field $\psi$ will have the following form
\begin{eqnarray}\label{virial}
V^\mu=\partial_\alpha\sigma^{\alpha\mu}=d\bigtriangleup\psi(x)\partial^\mu\psi+(d-4)\psi \partial^\mu\bigtriangleup\psi(x).
\end{eqnarray}
One can easily show that
\begin{eqnarray}\label{virial}
\partial_\mu V^\mu=d(\bigtriangleup\psi(x))^2+(2d-4)\partial_\mu\psi\partial^\mu\bigtriangleup\psi(x)
\end{eqnarray}
Using the above definitions  one can define the following traceless improved energy-momentum tensor \cite{CCJ} :
 \begin{eqnarray}\label{traceless}
T^t_{\mu\nu}=T^B_{\mu\nu}+\frac{1}{2}\partial_{\lambda}\partial_{\rho}X^{\lambda\rho\mu\nu},
\end{eqnarray}
where 
 \begin{eqnarray}\label{X}
X^{\lambda\rho\mu\nu}=\frac{2}{d-2}\Big{(}\delta^{\lambda\rho}\sigma^{\mu\nu}-\delta^{\lambda\mu}\sigma^{\rho\nu}-\delta^{\lambda\nu}\sigma^{\rho\mu}+\delta^{\mu\nu}\sigma^{\lambda\rho}+\frac{1}{d-1}(\delta^{\lambda\rho}\delta^{\mu\nu}-\delta^{\lambda\mu}\delta^{\rho\nu})\sigma^\alpha_\alpha\Big{)}.
\end{eqnarray}
Then one can also write 
 \begin{eqnarray}\label{D}
D^\mu=x^\nu T^{t\mu}_\nu.
\end{eqnarray}

The above procedure is not repeatable for $d = 2$ because $X^{\lambda\rho\mu\nu}$ in (\ref{X})
becomes singular. This is related to the fact that the action (\ref{fourth derivative scalar theory})
does not have full conformal symmetry, as we already know.  It is only invariant  under  M\"obius transformations. The situation is very similar to the non-local field theories that we have discussed in  sections ~2 and ~3.

\end{document}